\documentclass {PoS}
 \usepackage{epsf,epsfig}
\newcommand{\nn}{\nonumber}

\def\xb{\overline{x}}

\def\veps{\varepsilon}
\def\als{\alpha_s}
\def\vk{{\bf k}_{\perp}}

\def\gev{\,{\rm GeV}}
\title{Diffractive vector meson leptoproduction and spin effects}

\ShortTitle{Diffractive vector meson leptoproduction}

\author{S.V.Goloskokov\\
         Bogoliubov Laboratory of Theoretical
Physics, Joint Institute
for Nuclear Research,\\ Dubna 141980, Moscow region, Russia\\
        E-mail: \email{goloskkv@theor.jinr.ru}}

\abstract{We analyse spin effects in diffractive vector meson
leptoproduction at small $x$ on the basis of the generalized
parton distribution (GPD) approach. We take into account quark
transverse degrees of freedom in the hard subprocess. We calculate
amplitudes for the longitudinally and transversely polarized
photons and vector mesons. Our results on the cross section and
spin density matrix elements (SDME) are in fair agreement with the
DESY experiments. Predictions for HERMES and COMPASS energy range
are made. The predicted double spin longitudinal $A_{LL}$
asymmetry is not small at HERMES energies.}

\FullConference{DIFFRACTION 2006 - International Workshop on Diffraction in High-Energy Physics \\
         September 5-10 2006\\
         Adamantas, Milos island, Greece}

\begin{document}
 \section{Introduction}

In this report, we study spin effects in vector meson
leptoproduction at large energies within the GPD approach on the
basis of \cite{gk05}. At small $x$-Bjorken ($x$) the process
factorizes \cite{fact} into a hard meson leptoproduction off
gluons and GPD (Fig. 1).

\begin{figure}[h]
\centering \mbox{\epsfysize=30mm\epsffile{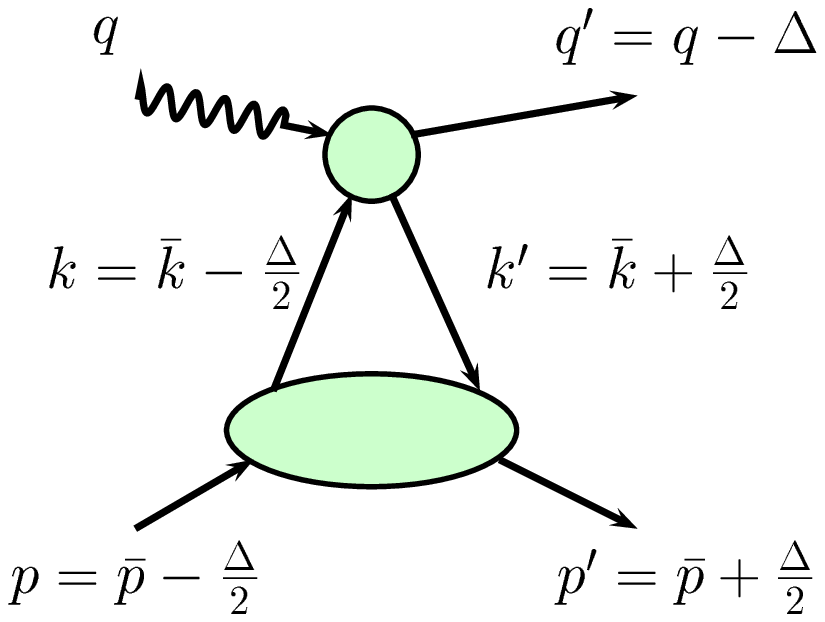}}
\caption{The handbag diagram for the meson electroproduction off
proton.} \label{kt_h}
\end{figure}

Unfortunately, it is difficult to study spin effects in  vector
meson leptoproduction  in the generally used standard collinear
approximation (SCA). Only the amplitude with longitudinally
polarized virtual photon and vector meson (LL) amplitude can be
calculated in the SCA. To analyse spin effects, one should have
information about ampitudes with transversally polarized photons.
The  TT and TL transition amplitudes where the  symbols $L$ and
$T$ refer to longitudinally and transversely polarized photons and
vector mesons, respectively, exhibit in the SCA the infrared
singularities \cite{mp} which breakdown  factorization.

Our model is based on the modified perturbative approach (MPA)
\cite{sterman} which includes the quark transverse degrees of
freedom accompanied by Sudakov suppressions. The transverse quark
momentum regularizes the end-point singularities in the TT and TL
amplitudes. This gives a possibility to calculate these amplitudes
in the MPA and suppress a contribution from the end-point region
in the LL amplitude. As a result, we obtain a reasonable
description of the H1 and ZEUS data \cite{h1,zeus}  for
electroproduced $\rho$ and $\phi$ mesons at small $x$ \cite{gk05}.

In this report, we  analyse  the $\phi$ meson leptoproduction.
Within the MPA we calculate  the three transition $LL$, $TT$ and
$TL$ amplitudes. We present our results for the cross sections and
SDME at DESY energies and give some predictions for the HERMES and
COMPASS energy range.

\section{Leptoproduction of  Vector Mesons in the GPD approach}
The leading-twist wave function  describes the longitudinally
polarized vector mesons. To study spin effects, we should consider
the higher order wave function which describes transversely
polarized mesons. Here we use the  $k$- dependent wave function
proposed in \cite{koerner}.

The gluon contribution to the leptoproduction amplitudes is
predominant in the low-$x$ region. For $t \sim 0$ and positive
proton helicity the amplitude reads as a convolution of the hard
subprocess amplitude ${\cal H}^V$ and a large distance unpolarized
$ H^g$ and polarized $\widetilde{H}^g$ gluon GPDs (Fig. 1):
\begin{eqnarray}\label{amptt-nf-ji}
  {\cal M}_{\mu'+,\mu +} &=& \frac{e}{2}\, {\cal C}_V\,
         \int_0^1 \frac{d\xb}
        {(\xb+\xi)
                  (\xb-\xi + i{\veps})}\nn\\
        &\times& \left\{\, \left[\,{\cal H}^V_{\mu'+,\mu +}\,
         + {\cal H}^V_{\mu'-,\mu -}\,\right]\,
                                   H^g(\xb,\xi,t) \right. \nn\\
       &+& \hspace{2.5mm}\left. \left[\,{\cal H}^V_{\mu'+,\mu +}\,
            -   {\cal H}^V_{\mu'-,\mu -}\,\right]\,
                        \widetilde{H}^g(\xb,\xi,t)\, \right\}\,.
\end{eqnarray}

Here $\mu$ ($\mu'$) denotes the helicity of the photon (meson),
$\xb$ is the  momentum fraction of the transversally polarized
gluons, and the skewness $\xi$ is related to Bjorken-$x$ by
$\xi\simeq x/2$. The flavor factors are $C_{\rho}=1/\sqrt{2}$ and
${ C}_{\phi}=-1/3$.  The polarized $\widetilde{H}^g$ GPD at small
$\xb$ is much smaller than the GPD $H^g$ and will be important in
the $A_{LL}$ asymmetry only.

 The subprocess amplitude ${\cal H}^V$ is represented as  the contraction of the hard
  part $F$, which is calculated perturbatively, and the
non-perturbative meson  wave function $ \phi_V$
\begin{equation}\label{hsaml}
  {\cal H}^V_{\mu'+,\mu +}\,\pm\,  {\cal H}^V_{\mu'-,\mu -}\,=
\,\frac{2\pi \als(\mu_R)}
           {\sqrt{2N_c}} \,\int_0^1 d\tau\,\int \frac{d^{\,2} \vk}{16\pi^3}
            \phi_{V}(\tau,k^2_\perp)\;
                  F_{\mu^\prime\mu}^\pm .
\end{equation}
  The wave function is chosen  in the Gaussian form
\begin{equation}\label{wave-l}
  \phi_V(\vk,\tau)\,=\, 8\pi^2\sqrt{2N_c}\, f_V a^2_V
       \, \exp{\left[-a^2_V\, \frac{\vk^{\,2}}{\tau\bar{\tau}}\right]}\,.
\end{equation}
 Here $\bar{\tau}=1-\tau$, $f_V$ is the
decay coupling constant and the $a_V$ parameter determines the
value of average transverse momentum of the quark in the vector
meson. Generally, values of $f_V, a_V$ are  different for
longitudinal and transverse polarization of the meson.

In calculation of the subprocess amplitude  within the MPA
\cite{sterman}  we keep the $k^2_\perp$ terms in the denominators
of the amplitudes. We keep  the $k^2_\perp$ terms in the numerator
of the TT amplitude too. The gluonic corrections are treated in
the form of the Sudakov factors which additionally suppress the
end-point integration regions. In the model the following
hierarchy of the amplitudes is obtained
\begin{equation}\label{hier}
{\rm LL}:\;\;\;  {\cal M}_{0\,\nu,0\,\nu}^{V(g)} \propto
1\,;\qquad {\rm TL}:\;\;\;  {\cal M}_{0\,\nu,+\nu}^{V(g)} \propto
                                                  \frac{\sqrt{-t}}{Q};\qquad
{\rm TT}:\;\;\;{\cal M}_{+\nu,+\nu}^{V(g)} \propto \frac{\vk^2}{Q
M_V}.
 \end{equation}
The  $L \to T$ and $T \to -T$ transition amplitudes are small and
neglected in our analysis.

The GPD is a complicated function which depends on three
variables. For a small momentum transfer  $t \sim 0$ we  use the
double distribution form \cite{mus99}
\begin{equation}
  H^{g}(\xb,\xi,t) = \Big[\,\Theta(0\leq \xb\leq \xi)
    \int_{\frac{\xb-\xi}{1+\xi}}^{\frac{\xb+\xi}{1+\xi}}\, d\beta +
     \Theta(\xi\leq \xb\leq 1) \int_{\frac{\xb-\xi}{1-\xi}}
     ^{\frac{\xb+\xi}{1+\xi}}\, d\beta \,\Big]\,
 \frac{\beta}{\xi}\,f(\beta,\alpha=\frac{\xb-\beta}{\xi},t)
\end{equation}
with the simple factorizing ansatz for the double distributions
$f(\beta,\alpha,t)$
\begin{equation}
  f(\beta,\alpha,t\simeq 0) = g(\beta)\,\frac{3}{4}\,
\frac{[(1-|\beta|)^2-\alpha^2]}{(1-|\beta|)^{3}}.
\end{equation}
The ordinary gluon distribution is parameterized as
\begin{equation}\label{gd}
\beta g(\beta)= \beta^{-\delta(Q^2)}\,(1-\beta)^5\, \sum_{i=0}^2
c_i\, \beta^{\,i/2}.
\end{equation}
The parameter $\delta(Q^2)$
\begin{equation}\label{delta}
\delta(Q^2) = 0.17 + 0.07 \ln{(Q^2/Q_0^2)} - 0.005
\ln^2{(Q^2/Q_0^2)}
\end{equation}
determines the  behavior of the gluon distribution at low $x$ and
the energy dependence of the cross section at high energies. Other
parameters in (\ref{gd}) are taken from comparison with the CTEQ5M
parameterization ~\cite{CTEQ}.

\begin{figure}[h]
\centering \mbox{\epsfysize=50mm\epsffile{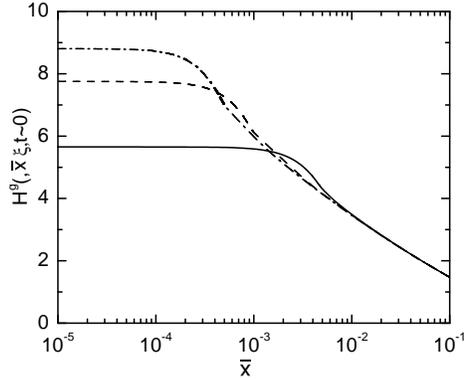}}
\caption{Model results for the GPD $H^g$ in the small $\xb$ range.
The solid (dashed,  dash-dotted) line represents the GPD at $\xi=
5\;(1\,,\; 0.5)\, \cdot 10^{-3}$}
\end{figure}
This model gives a possibility to calculate GPD \cite{gk05}
through the gluon distribution $g(\beta)$.  For the three $\xi$
values the gluon GPD is shown in Fig. 2.

\section{Cross section and spin observables}
Experimentally, only the slope of the $\gamma^* p\to Vp$ cross
section is measured. The  $t$- dependence of the amplitudes is
generally unknown. For simplicity, we parameterize it in the
exponential form $M_{ii}(t)=M_{ii}(0)\; e^{t\,B_{ii}/2}$ for $
LL,\; TT,\; TL$ transitions and suppose   that  $ B_{LL}\sim
B_{TT} \sim B_{TL}$.

\begin{figure}[h!]
\begin{center}
\begin{tabular}{cc}
\mbox{\epsfig{figure=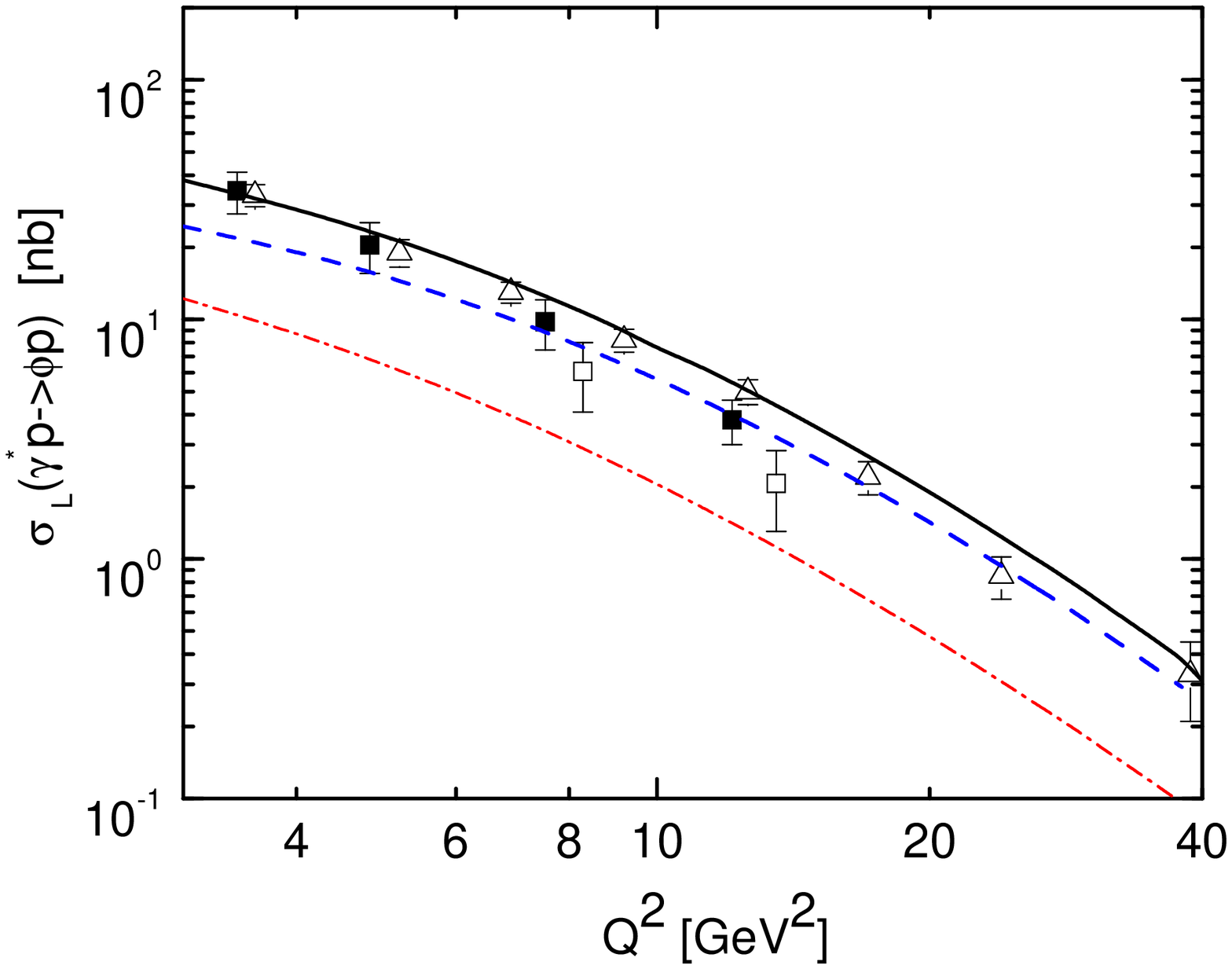,width=6.2cm,height=5cm}}&
\mbox{\epsfig{figure=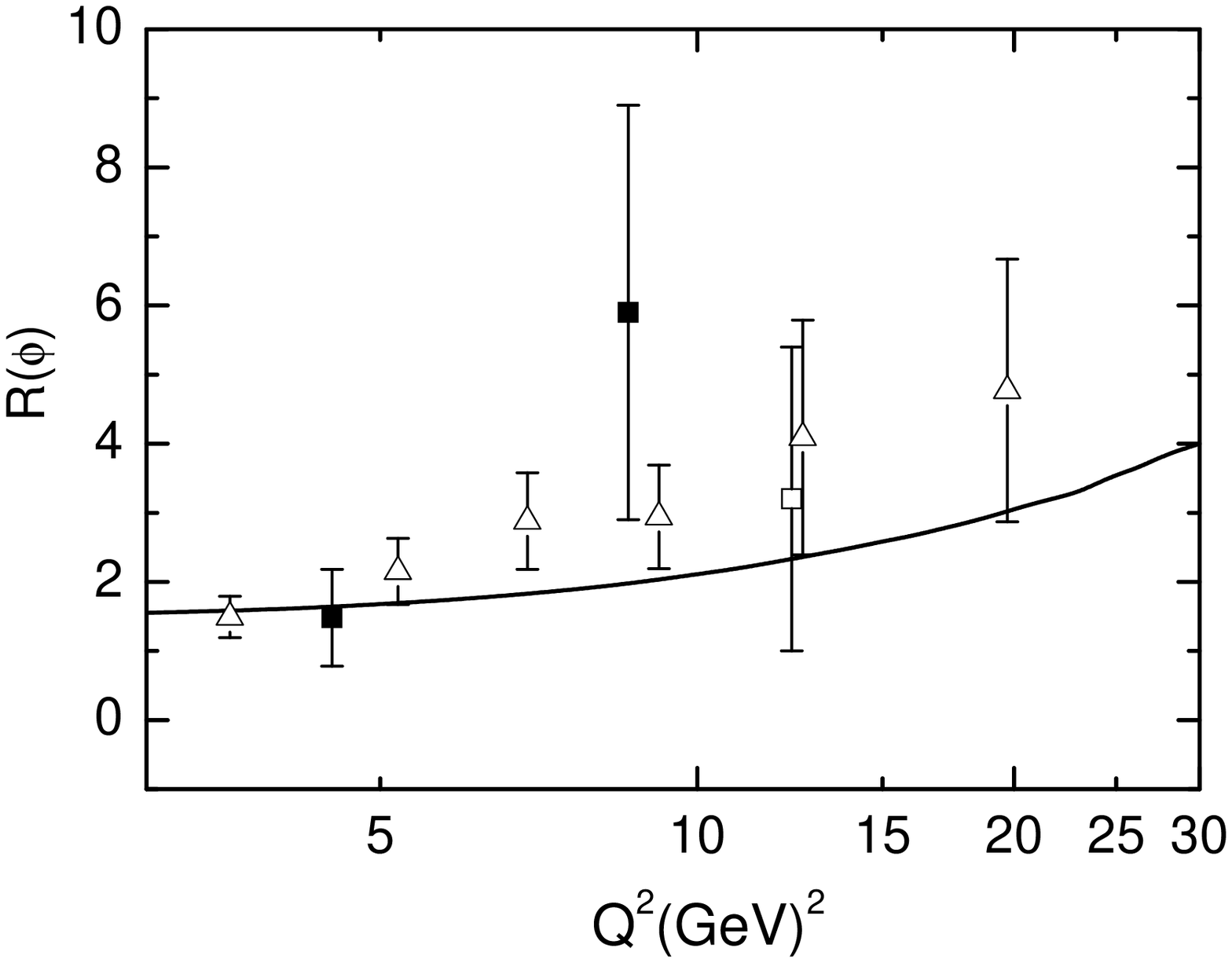,width=5.8cm,height=5cm}}\\
{\bf(a)}& {\bf(b)}
\end{tabular}
\end{center}
\caption{ {\bf(a)} Longitudinal cross sections of $\phi$
production at $W=75 \mbox{GeV}$. Full line cross section, dashed-
gluon contribution, dashed-dot - gluon-strange interference.
{\bf(b)} The ratio of longitudinal and transverse cross sections
 for the $\phi$  production  at
 $W =75\, \gev$ and $t=-0.15\,\gev^2$. Data are  from H1 and ZEUS.}
\end{figure}

The longitudinal cross section for the $\phi$
 production integrated over $t$  is shown
in Fig. 3a.  Estimations for the vector meson production
 are obtained using $f_{\phi L}=.237\gev$, $a_{\phi
L}=0.45\gev^{-1}$; $f_{\phi T}=.190\gev$; $a_{\phi
T}=0.6\gev^{-1}$. In addition to the gluon distribution the
gluon-sea-quark interference is shown in   Fig. 3a. It can be seen
that the typical contribution of the  interference to $\sigma_L$
does not exceed 40 \% with respect to the gluon one. Thus, the
gluon term really gives the predominant contribution to the cross
section and we find  good agreement of our results with H1 and
ZEUS experiments \cite{h1} at HERA. Note that the uncertainties in
the gluon GPD provide errors in the cross section about $25-35
\%$. This is the same order of magnitude as the gluon-sea
interference. The model results for the ratio of the cross section
with longitudinal and transverse photon polarization $R$ for the
$\phi$ production are shown in Fig. 3b and are  consistent with
the H1 and ZEUS experiments \cite{h1,zeus}.

\begin{figure}[h!]
\begin{center}
\begin{tabular}{ccc}
\mbox{\epsfig{figure=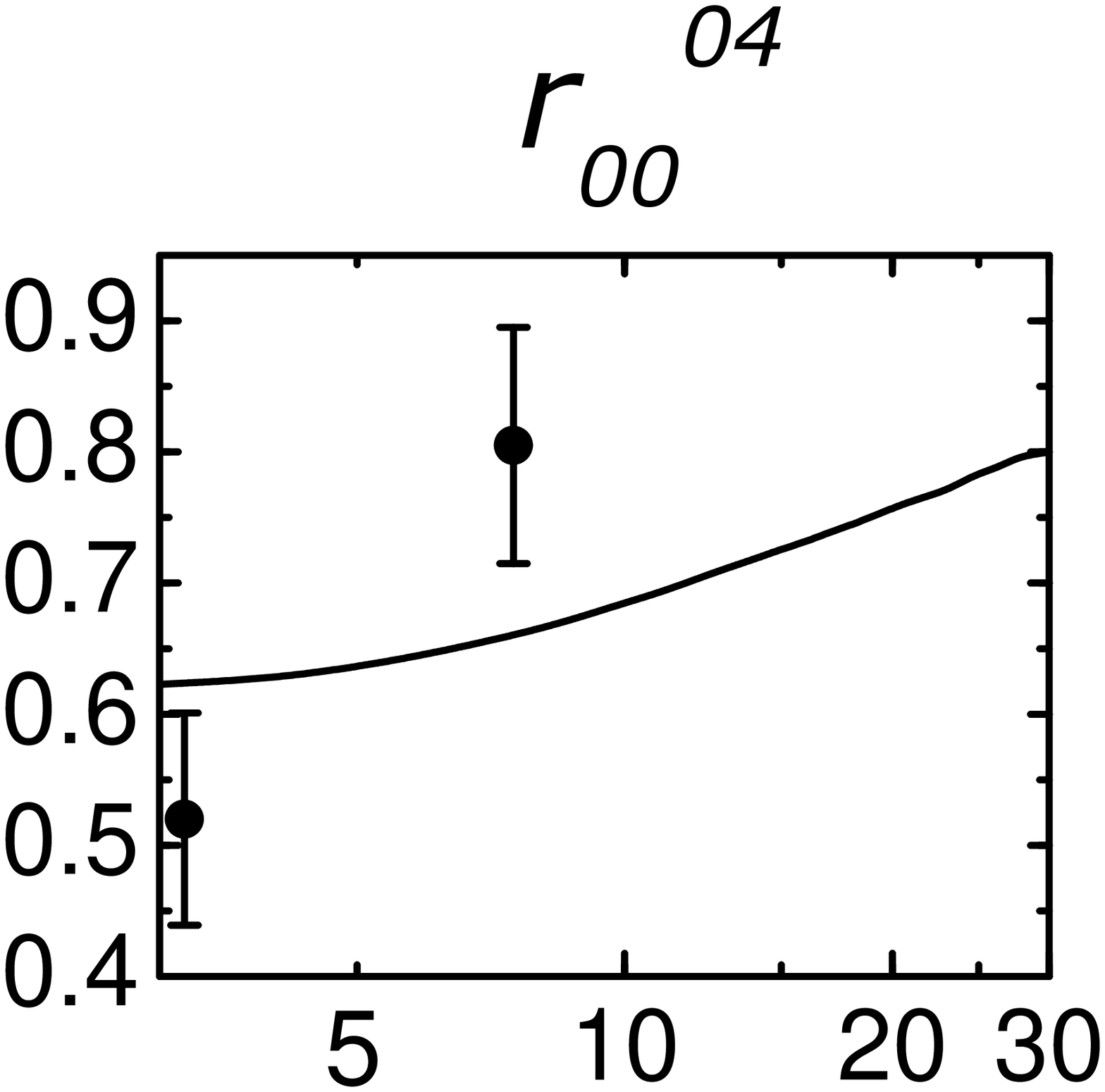,width=3.4cm,height=3.cm}}&
\mbox{\epsfig{figure=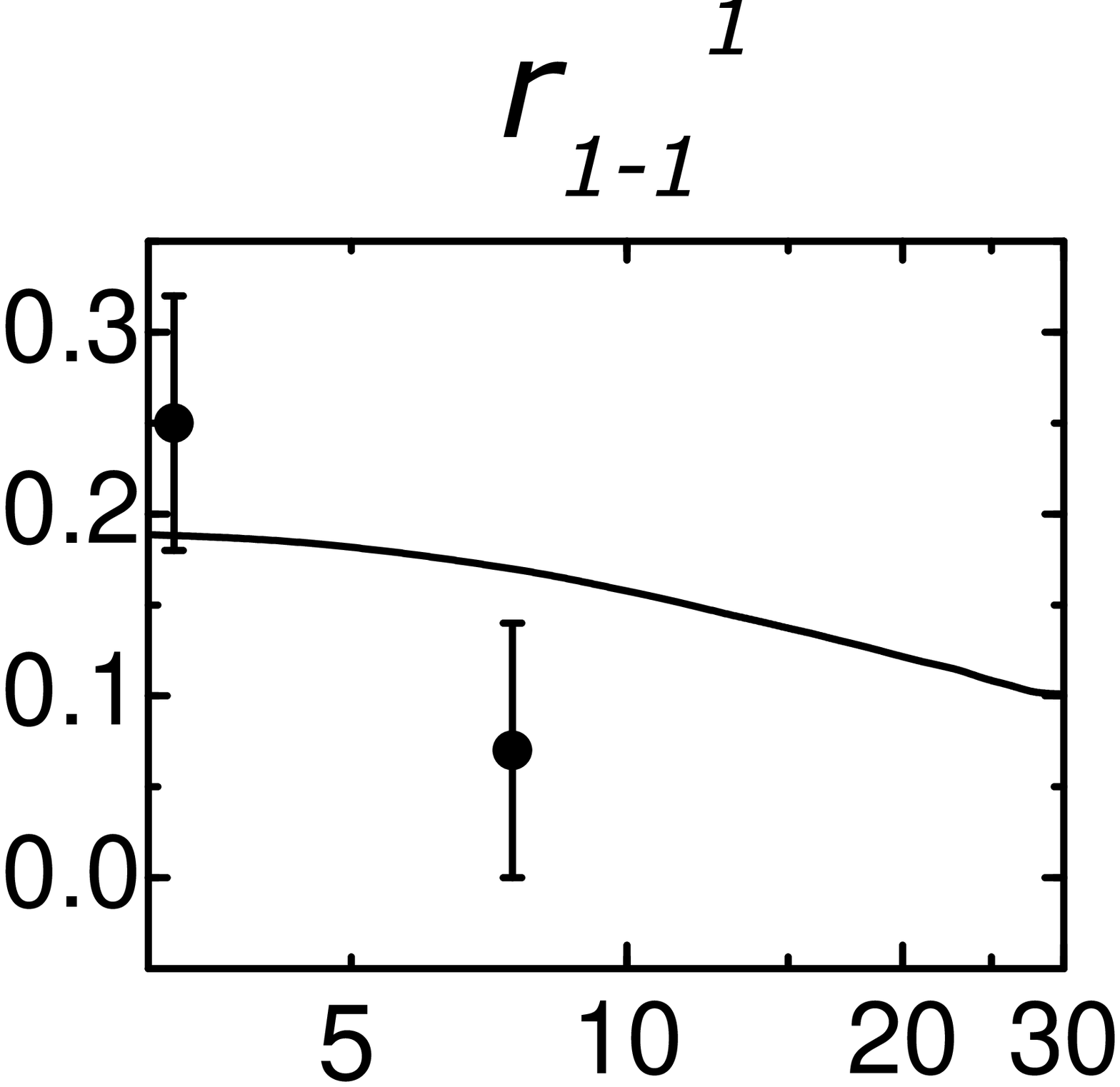,width=3.4cm,height=3.cm}}&
\mbox{\epsfig{figure=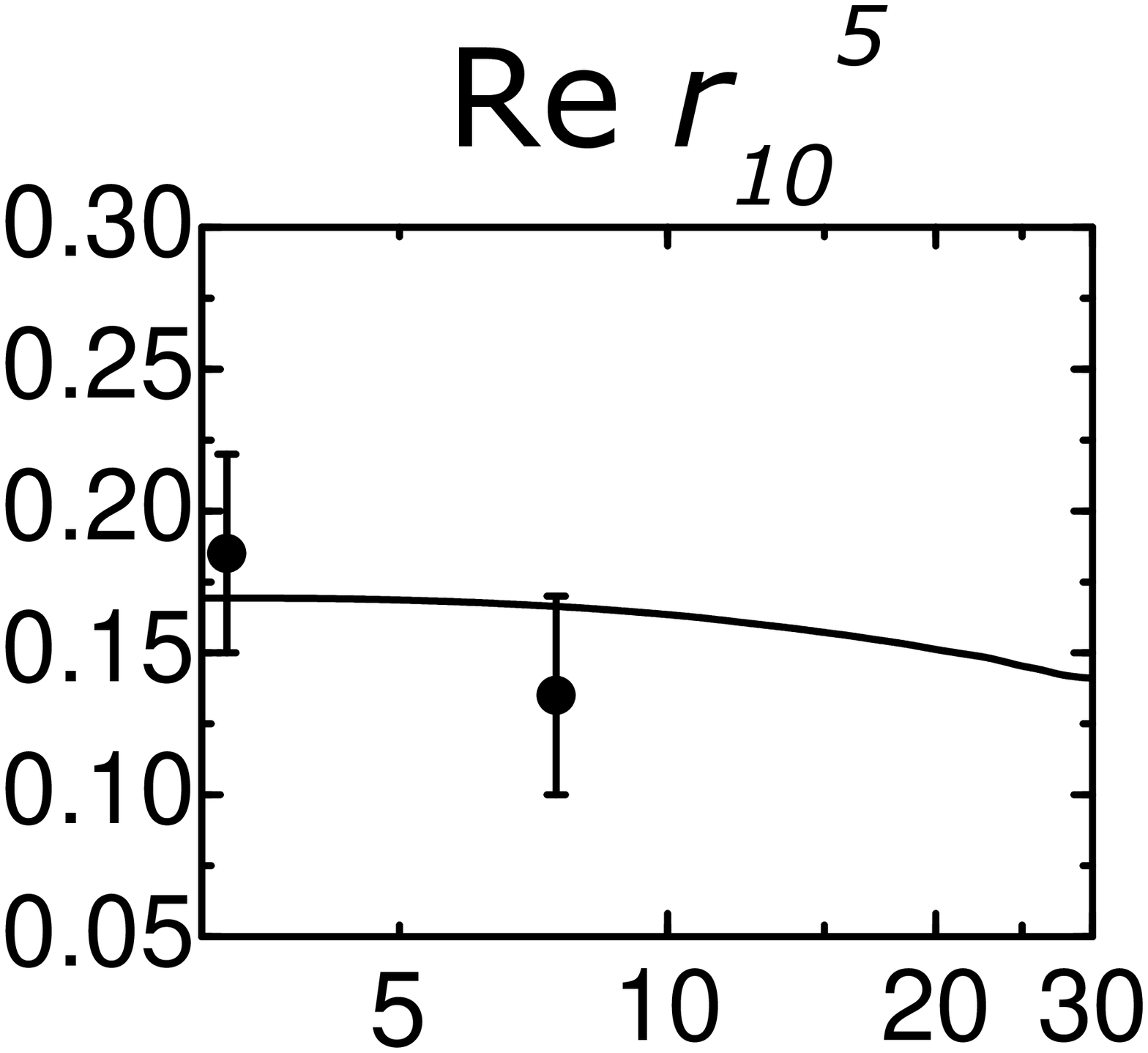,width=3.4cm,height=3.cm}}\\
\end{tabular}
\phantom{aa}\vspace{-1mm} \centerline{$Q^2[\mbox{GeV}]^2$}
\end{center}
\caption{ The $Q^2$ dependence of SDME on the $\phi$ production at
$t=-.15\gev^2$ and $W=75\gev$.   Data are taken from H1. }
\end{figure}

SDME are determined in terms of interference of different
amplitudes of vector meson photoproduction. Thus, they are
sensitive to the LL,TT, TL amplitude structure and to the vector
meson wave function. In Fig. 4, we present our results for the
three SDME in the DESY energy range. Description of experimental
data \cite{h1} is reasonable. Our results for other SDME can be
found in \cite{gk05}. We can conclude that our amplitude hierarchy
(\ref{hier}) is supported by experiment.

Let us discuss the energy dependence of the cross section. At
small $x$ they behave as
\begin{equation}\label{sigma}
\sigma_L \propto W^{4 \delta(Q^2)}.
\end{equation}
In the Regge picture  the power $\delta$  (\ref{delta}) is
associated with the Pomeron intercept $\alpha_P(0)=1+\delta(Q^2)$.
In Fig. 5, we show our results for the $\phi$- production cross
section. The data \cite{zeus} in Fig. 5 are  recalculated  to the
longitudinal cross section using the ratio  $R= 1.5$ at $Q^2=3.8
\mbox{GeV}^2$ \cite{zeus}. One can see that the preliminary HERMES
\cite{bian} and HERA data \cite{h1,zeus} are described well.
\begin{figure}[h]
\centering \mbox{\epsfysize=46mm\epsffile{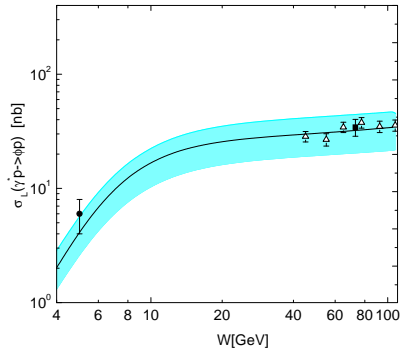}}
\caption{The longitudinal cross section for $\phi$ production vs.
$W$ for $Q^2=3.8\, \mbox{GeV}^2$.}
\end{figure}

Thus, we can make some predictions for $\phi$ production SDME at
intermediate energies. Our results for COMPASS are shown in Fig.
6. It can be seen that the energy dependencies of SDME are weak.

\begin{figure}[h!]
\begin{center}
\begin{tabular}{ccc}
\mbox{\epsfig{figure=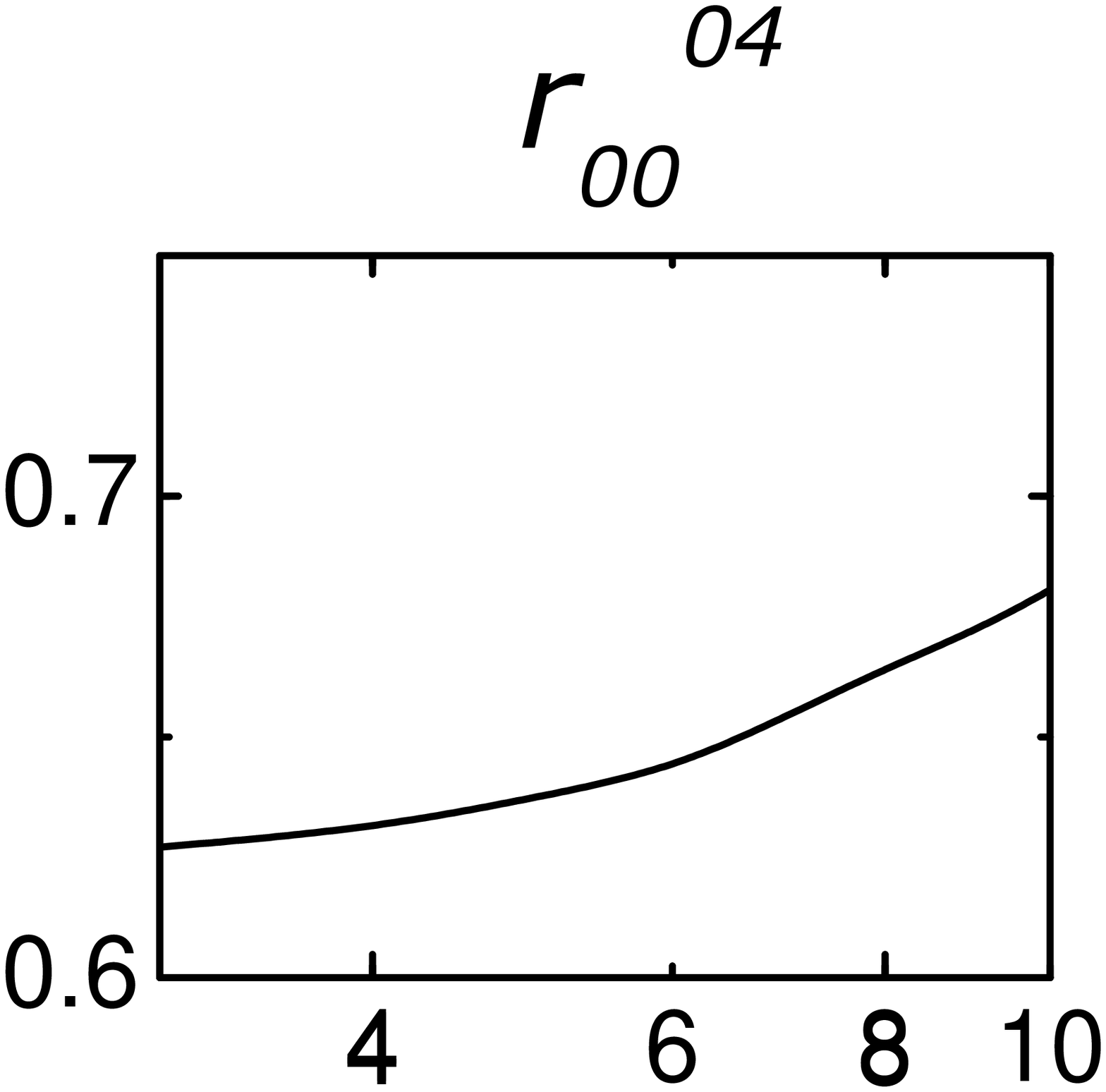,width=3.4cm,height=3.cm}}&
\mbox{\epsfig{figure=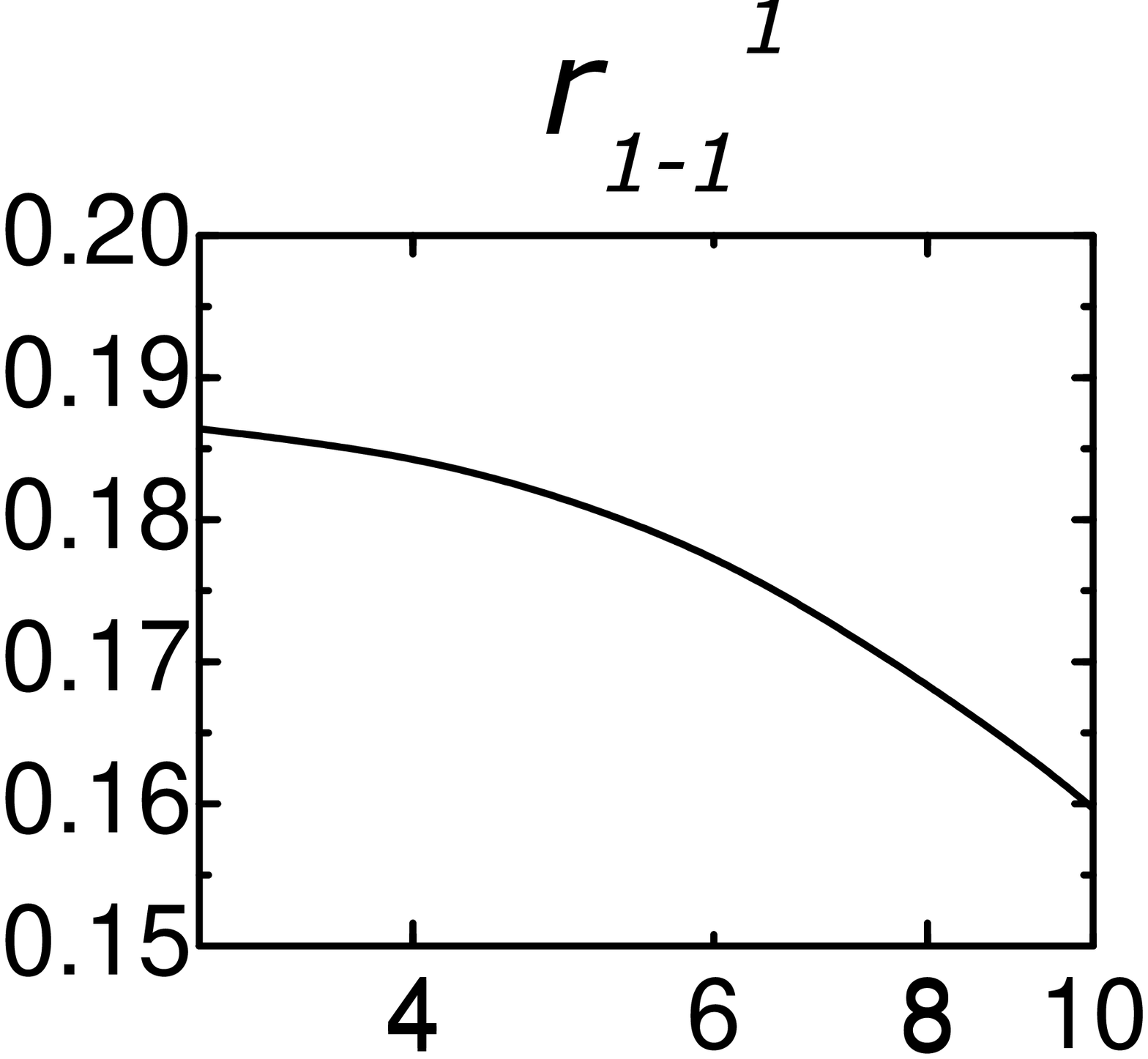,width=3.4cm,height=3.cm}}&
\mbox{\epsfig{figure=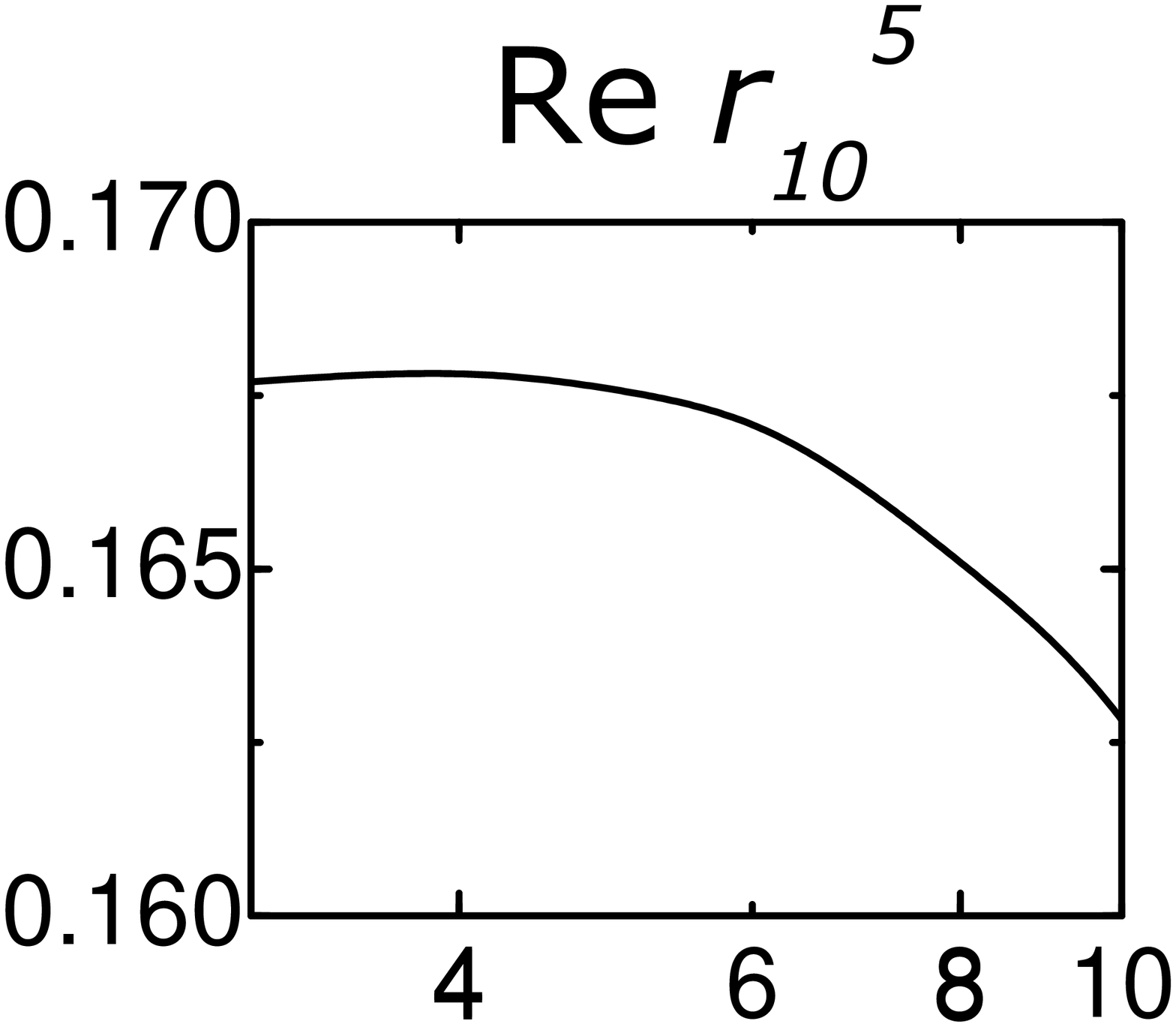,width=3.4cm,height=3.cm}}\\
\end{tabular}
\phantom{aa}\vspace{-1mm} \centerline{$Q^2[\mbox{GeV}]^2$}
\end{center}
\caption{ The predicted $Q^2$ dependence of SDME on the $\phi$
production at $t=-.15\gev^2$ and the COMPASS energy $W=10\gev$. }
\end{figure}

The $A_{LL}$ asymmetry for a longitudinally polarized beam and
target is sensitive to the polarized GPD.  Really, the leading
term in the $A_{LL}$ asymmetry integrated over the azimuthal angle
is determined through the interference between the unpolarized GPD
$H^g$ and  polarized $\widetilde{H}^g$ distributions.

We expect a small  value for the $A_{LL}$ asymmetry at high
energies because it is of the order of the ratio $\langle
\widetilde{H}^g\rangle/\langle H^g \rangle$ which is small  at low
$x$.
\begin{figure}[h]
\begin{center}
\mbox{\epsfig{figure=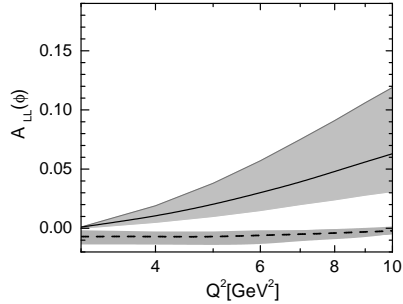,width=5.2cm,height=4cm}}
\end{center}
\caption{ The $A_{LL}$ asymmetry of $\phi$ meson production at
$W=5\,\gev$ (solid line) and
  $W=10\,\gev$ (dashed line); $y\simeq 0.6$. The shaded bands reflect
  the uncertainties in $A_{LL}$ due to the  error in
  $\widetilde{H}^g \propto \Delta g$.}
  \label{yourname_fig4}
\end{figure}
At COMPASS energies $W=10 \mbox{GeV}$ the gluon contribution to
the $A_{LL}$ asymmetry of vector meson  production is  small (Fig.
7). At HERMES energies $W=5 \mbox{GeV}$ the major contribution to
the amplitudes comes from the region $0.1 \le \xb \le 0.2$ where
$\Delta g/g$ is not small, which leads to a large value of the
$A_{LL}$. The asymmetry is shown in Fig. 7 together with the
uncertainties in our predictions due to the error in the polarized
gluon distribution. Note that our predictions should be  valid for
the $\phi$ production. In case of $\rho$ production the polarized
quark GPD should be essential in the $A_{LL}$ asymmetry in the low
energy range.

 \section{Conclusion or Summary}
We have analyzed electroproduction of $\phi$  meson at small
Bjorken-$x$ in the handbag approach  where the process amplitudes
are factorized into the gluonic generalized parton distributions
and a partonic subprocess, $\gamma g \to (q\bar{q}) g$,
$q\bar{q}\to V$. The subprocess was calculated \cite{gk05} within
the modified perturbative approach where the transverse momenta of
the quark and antiquark as well as Sudakov corrections were taken
into account. These effects suppress the contributions from the
end-point regions where one of the partons entering the meson wave
function becomes soft and factorization breaks down in the
collinear approximation. This allowed us to investigate the
amplitudes for all kinds of transitions from longitudinally and
transversally polarized virtual photons to polarized vector
mesons.
 This give a possibility to study spin effects in vector meson production.
Within the GPD approach we find fine description of the $Q^2$
dependence of the cross section, $R$ ratio and   SDME for the
$\rho$ meson production at HERA energies. More results for $\rho$
and $\phi$ production can be found in \cite{gk05}. Some comparison
with the two-gluon exchange model \cite{bro94}  can be found in
\cite{gk05} too.

 We would like to point out that study of SDME gives an important
 information about different  $\gamma \to V$ hard amplitudes. By analyses of $t$
dependences of SDMEs  the diffraction peak slopes of the $TT$ and
$TL$ amplitudes can be defined. Unfortunately, the data on spin
observables have  large experimental errors. The new experimental
results with reduced errors in SDME at $Q^2\geq 3\mbox{GeV}^2$ are
extremely important. Information about polarized gluon
distribution can be obtained from the $A_{LL}$ asymmetry  on
$\phi$ meson production   which is sensitive to $\Delta G$. Thus,
we can conclude that the vector meson photoproduction at small $x$
is an excellent tool  to study the gluon  GPD.

 \bigskip

 This work is supported  in part by the Russian Foundation for
Basic Research, Grant 06-02-16215  and by the Heisenberg-Landau
program.

\end{document}